\documentclass{iaus}
\usepackage{graphicx}
\usepackage{graphics}

\title[BSP Synthesis Model] 
{Binary Stellar Population Synthesis Model}

\author[F. Zhang, Z. Han \& L. Li]   
{F. Zhang$^1$,
 Z. Han$^1$, \and L. Li$^1$}

\affiliation{$^1$ National Astronomical Observatories/Yunnan Observatory,
Chinese Academy of Sciences, P.O. Box 110, Kunming, Yunnan Province 650011,
China \break email: zhang\_fh@hotmail.com \\[\affilskip]
}

\pubyear{2006}
\volume{241}  
\pagerange{xxx--xxx}
\date{?? and in revised form ??}
\jname{Stellar Populations as Building Blocks of Galaxies}
\editors{xxx Editor, eds.}
\begin{document}

\maketitle

\begin{abstract}
Using Yunnan evolutionary population synthesis (EPS) models, we
present integrated colours, integrated spectral energy
distributions (ISEDs) and absorption-line indices defined by the
Lick Observatory image dissector scanner (Lick/IDS) system, for an
extensive set of instantaneous-burst binary stellar populations
(BSPs) with interactions. By comparing the results for populations
with and without interactions we show that {\it the inclusion of
binary interactions makes the appearance of the population
substantially bluer}. This effect raises the derived age and
metallicity of the population.

To be used in the studies of modern spectroscopic galaxy surveys
at intermediate/high spectral resolution, we also present
intermediate- (3\,\AA) and high-resolution ($\sim 0.3$\,\AA) ISEDs
and Lick/IDS absorption-line indices for BSPs. To directly compare
with observations the Lick/IDS absorption indices are also
presented by measuring them directly from the ISEDs.

\keywords{binaries: general, stars: evolution, galaxies: clusters:
general}
\end{abstract}

Binary stars play an important role in the evolution of stellar
population. First, observations show that upwards of 50 per cent
of the stars populating galaxies are expected to be in binary or
higher-order multiple systems. Secondly, in some binary systems
the evolution of the component stars are significantly different
from those expected from single-star evolution. The inclusion of
binaries would alter the overall appearance of the population.
However, binary interactions have been neglected in some
evolutionary population synthesis (EPS) studies. In our studies we
assume that all stars ($2 \times 10^6$) are born in binaries and
born at the same time, i.e., an instantaneous binary stellar
population (BSP), then model any interactions within these
binaries in our Yunnan EPS models.

{\bf MODELS:} A full description of Yunnan EPS model and algorithm
are given in \cite[Zhang et al.
(2004,2005a,b)]{zha04,zha05a,zha05b}, we refer the interested
reader for them. For $Z=0.02$ standard model $\sim\,11.6\,\%$ of
the binaries would experience Roche lobe overflow (RLOF) during
the past 13 Gyr.

{\bf RESULTS:} We present the integrated colours, Lick/IDS
absorption-line indices and integrated spectral energy
distributions (ISEDs) for BSPs at low (10-20\,\AA), intermediate
(3\,\AA) and high ($\sim 0.3$\,\AA) resolutions. The
low-resolution ISEDs cover the range 91-1600,000\,\AA,
intermediate- and high-resolution ISEDs in the range of
3000-7000\,\AA. The Lick/IDS absorption-line indices include 21
indices of \cite[Worthey et al. (1994)]{wor94}, four Balmer
indices defined by \cite{wor97} and 13 refined indices of
\cite[Trager et al. (1998)]{tra98}. The synthetic BSP Lick/IDS
absorption indices are also measured directly from the
intermediate- and high-resolution ISEDs, therefore, it becomes to
be possible that compare theoretical results directly with
observations. The ages of BSPs are in the range 1-15\,Gyr, the
metallicity are in the range $0.0001-0.03$ at low resolution and
$0.004-0.03$ at intermediate and high resolutions.

\begin{figure}
\centering{
 \includegraphics[angle=270,scale=.20]{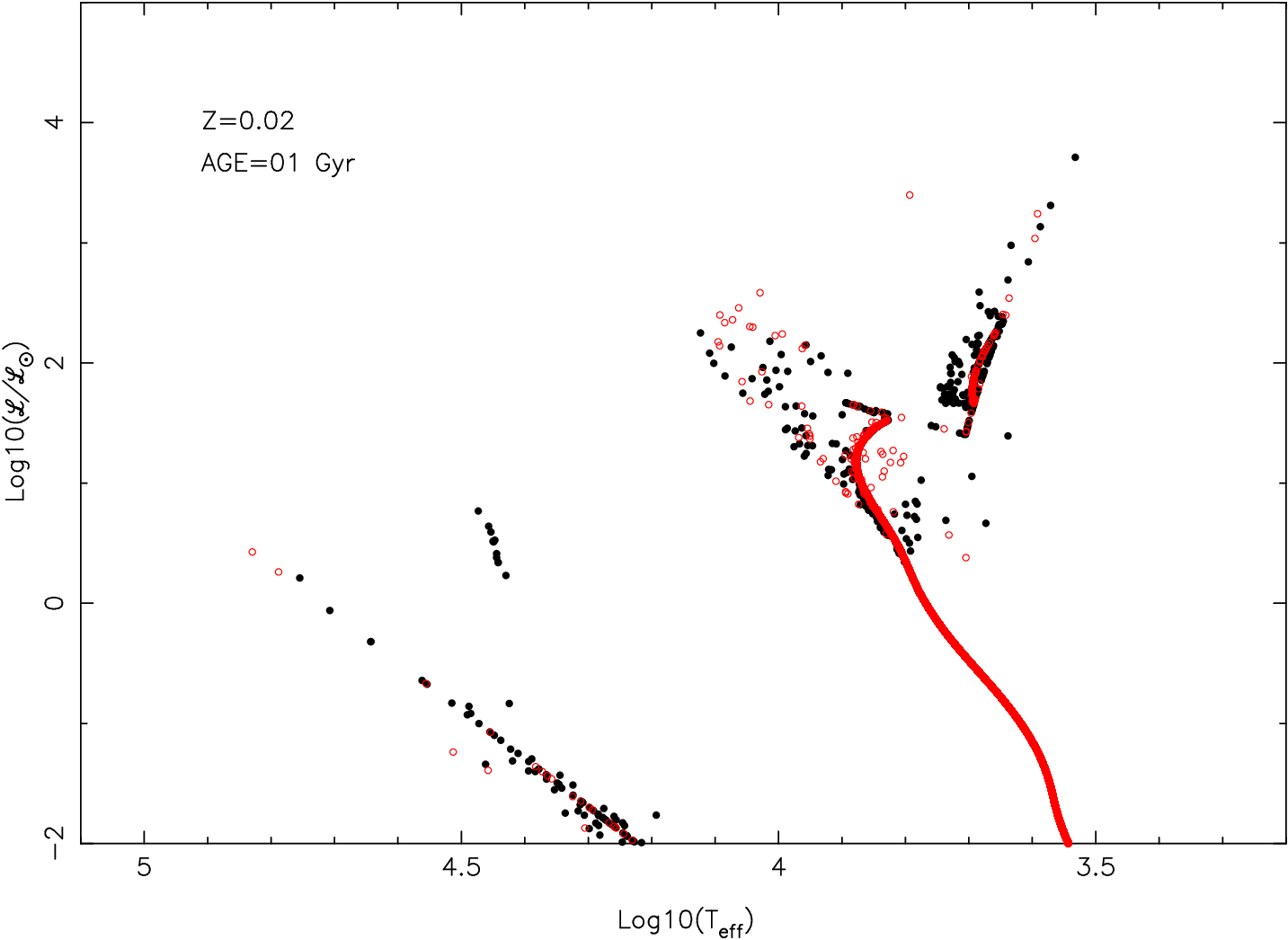}
 \includegraphics[angle=270,scale=.20]{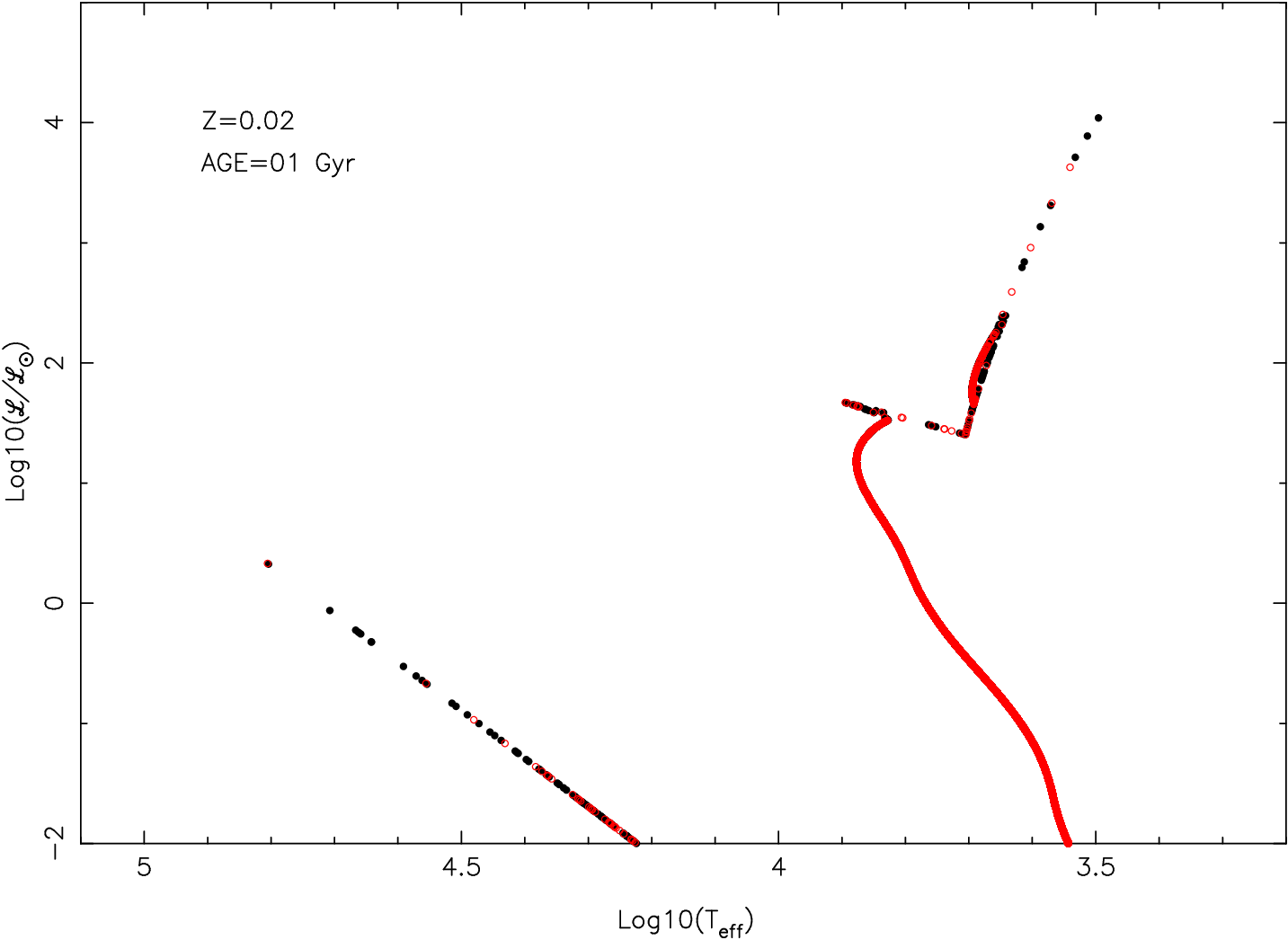}
 }
\caption{~Theoretical isochrones for $Z=0.02$ instantaneous burst
BSPs at an age of 1 Gyr (solid, red circles for the primary, open,
black for the secondary). The left and right panels represent
those considering and neglecting binary interactions,
respectively. For the sake of clarity only $5 \times 10^4$ binary
systems are included in each panel.}
  \label{fig01}
\end{figure}

In Fig. \ref{fig01} we present the theoretical isochrones of $\tau
=1$\,Gyr, $Z=0.02$ instantaneous BSPs with and without binary
interactions. It is seen clearly that the distribution of stars in
Hertzsprung-Russell diagram is significantly different. These
differences in the distribution of stars are responsible for
differences in the appearance of EPS models.

\begin{figure}
\centering
\includegraphics[angle=270,scale=.20]{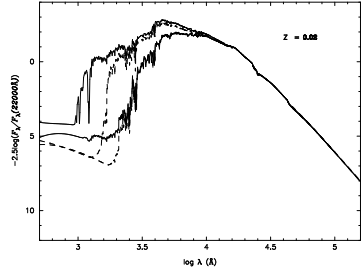}
\caption{~The ISEDs for solar-metallicity instantaneous burst BSPs
with (full line) and without (dashed line) binary interactions at
ages of 1 Gyr and 13 Gyr (upper and lower lines, respectively.}
\label{fig02}
\end{figure}

In Fig. \ref{fig02} we give a comparison of the low-resolution
ISEDs for $Z=0.02$ BSPs with and without binary interactions at
ages of 1\,Gyr and 13\,Gyr. We see that at $\tau=1$\,Gyr the ISED
of BSPs with binary interactions is as much as $\sim 5$ mag, at
$\tau=13$\,Gyr as much as $\sim 2$ mag greater than the
corresponding one without binary interactions in the
far-ultraviolet region. In the visible and infra-red regions the
BSPs with binary interactions exhibits bluer continuum than that
those without binary interactions. This causes $U-B$, $B-V$, $V-R$
and $R-I$ colours and Lick/IDS feature indices bluer by an factor
of 20\%, further, raises the derived age and metallicity of
populations \cite[(Zhang et al. 2004, 2005a)]{zha04,zha05a}.

To satisfy the needs of modern spectroscopic galaxy survey, we
also present the ISEDs and Lick/IDS absorption indices for BSPs at
intermediate (3\,\AA) and high ($\sim$ 0.3\,\AA) resolutions
\cite[(Zhang et al. 2005b, 2006)]{zha05b,zha06}.

This work was in part supported by Chinese Natural Science
Foundation (Grant Nos 10303006, 10673029, 10433030, 10521001),
Yunnan Natural Science Foundation (Grant No. 2005A0035Q) and the
Chinese Academy of Sciences (KJCX2-SW-T06).

\end{document}